\documentclass[aps,prl,twocolumn,superscriptaddress]{revtex4-1}

\usepackage{amsfonts}
\usepackage{amsmath}
\usepackage{amssymb}
\usepackage{graphicx}
\usepackage{float}
\usepackage[dvipsnames]{xcolor}
\begin{document}

\title{A facilitation-induced fluidization transition in supercooled water triggered by a few active molecules.}

\date{{\color{black}\today} }

\author{Quoc Tuan Truong}
\affiliation{Laboratoire de Photonique d'Angers EA 4464, Universit\' e d'Angers, Physics Department,  2 Bd Lavoisier, 49045 Angers, France.}

\author{Victor Teboul}
\email{victor.teboul@univ-angers.fr}
\affiliation{Laboratoire de Photonique d'Angers EA 4464, Universit\' e d'Angers, Physics Department,  2 Bd Lavoisier, 49045 Angers, France.}


\begin{abstract}
Using an activation mechanism reproducing facilitation, a dynamic phase transition triggered by a few active molecules was recently found in a supercooled model liquid.
Prompted by this finding we investigate the presence of a similar transition in supercooled water. 
We find the presence of the phase transition in water despite the numerous anomalies of water, suggesting  universality of the transition. 
The transition appears at constant temperature, being only induced by the motion of a small percentage of active molecules and the system's cooperativity. We observe that cooperative motions strongly increase at the transition and do not disappear when the medium's viscosity drops. 
An increase of temperature is needed to make the cooperative motions disappear, suggesting a connection to the kinetic energy to potential energy ratio instead of the medium's viscosity.

\end{abstract}

\maketitle
{
\section{ Introduction}

Media containing self-propelled particles, such as molecular motors or nanomachines, are classified as active matter\cite{active1,active2,active3,active4,active5,active6,active7,active8,active9,active10,motoro1,motoro2,motoro3,motoro4,motoro5,Szamel1,Szamel2,Szamel3,Szamel4,Szamel5,Szamel6,Szamel7}. This field has attracted considerable interest due to its relevance to biology and out-of-equilibrium statistical physics. Active matter also offers a novel framework for addressing the long standing glass transition problem\cite{gt0,gt1,gt2,anderson,fragile1}, owing to its non-equilibrium properties and potential underlying mechanisms.
Studies on supercooled active matter have demonstrated that the addition of active particles to a viscous medium reduces its viscosity without significantly increasing temperature, an effect termed fluidization.

A facilitation mechanism\cite{facile,facile1,facile2,facile3,facile5,facile6}, molecules motion being facilitated around already moving molecules, has been proposed\cite{facile,facile1} as the origin of the strange physics behind the glass transition.
Using an activation mechanism chosen to reproduce the facilitation, in recent work\cite{a1,a2} we studied the effect of a few activated molecules diluted inside a model\cite{ariane} supercooled liquid. 
We found a dynamic phase transition\cite{a1,a2} between the viscous liquid and a fluidized version of it retaining cooperative motions\cite{dh0,dh1,dh2,dh3,iso1,iso2,silica,prefold}.
Searching if the transition also appears in real liquids and the possible discrepancies due to their complexity we extend here our previous work to supercooled water\cite{wat0b,wat0,wat1,wat2,wat3,wat0a}. 

Water is a very particular liquid, displaying a number of physical anomalies\cite{wat3,wat1}. Therefore finding the same transition in water will suggest an underlying universal mechanism. Also water is the biological fluid, adding interest in this study and the prospect of  new applications.
Despite water's singularities we actually find in this work a similar fluidization transition in supercooled water than in our model system. Our results show that as in the model system study\cite{a1,a2}, only a small percentage of active molecules is necessary to trigger the fluidization transition. 
At a temperature of $220K$ only $2.5\%$ of active molecules are enough to trigger the transition, a percentage that decreases when the temperature increases.
It shows that only a small energy is involved to trigger the transition. 
Notice also the absence of thermal change during the transition, the temperature being controlled by a thermostat.

As in the model system we observe in supercooled water a large increase of cooperative mechanisms at the transition.
Cooperative mechanisms are usual signatures of phase transitions\cite{book-phystat1,book-phystat2}.
However the fact that these are the same cooperative mechanisms than spontaneous mechanisms observed in supercooled liquids,
suggests that they are the relevant cooperative mechanisms for the fluidization transition and  the glass transition.
The cooperative mechanisms do not disappear when the medium viscosity drops, but disappear when the temperature increases. 
We interpret that result as a sign that the cooperative motions are related to the kinetic energy, potential energy ratio, and survive temporary cage breaking mechanisms.
As found previously in the model system, our results also suggest in water a transmission of the cooperative motions from actives molecules to the non-actives, for the transition to occur.
Finally in contrast to previous results, we do not find a clear aggregation of active molecules in water, a result that we interpret as due to the interplay with a transition between LDL and HDL, also induced by activation. 
Notice that this other transition is of importance in the search to hinder water crystallization as LDL is known as a precursor to crystallization.

\section{Calculation}

A variety of intermolecular potentials exist to model water\cite{wpot1,wpot2,wpot3,wpot4,wpot5,wpot6,wpot7,wpot8,wpot8b,wpot8c,wpot8d}, sometimes with coarse graining for applications in biology. In this paper, we model the water molecular interactions with the TIP5PE potential\cite{wpot4,wpot5,wpot6} and the long range electrostatic interaction with the reaction field method\cite{md1,md2,md3} using a cutoff radius $R_{RF} = 9 $\AA\ and an infinite dielectric constant for distances $r > R_{RF}$. The water molecule is modeled as a rigid body, and we will focus our attention on the center of mass behavior (that is also approximately the oxygen atom behavior as the differences in the position of the center of mass and of the oxygen atom are quite small in water).
We solve the equations of motion using the Gear 4 algorithm\cite{md1,md2,md3} with a time step $\Delta t=10^{-15} s$.

A  thermostat\cite{berendsen}  removes the energy dissipated into the system avoiding a  drift in energy. 
In our calculations,  a concentration $C$ of the medium molecules (typically a few percents) are activated (pushed) in the mobility direction of the most mobile of their neighbors during $10 ps$ then the activation stops during $30ps$ leading to a  $40 ps$ periodicity. Therefore, an average of $C/4$ molecules are activated, while the other molecules do not experience external forces but only the intermolecular interactions. 

In this study, we investigate the alteration of key features in supercooled liquids, such as dynamic heterogeneity, diffusion properties, and the $\alpha$ relaxation time, which is associated with the viscosity of the medium. We will now define the statistical functions utilized for this purpose. One function of significant relevance in glass-transition phenomena is the intermediate scattering function, denoted as $F_{S}(Q,t)$, which portrays the autocorrelation of density fluctuations at the wave vector $Q$. This function provides insights into the structural relaxation of the material. We define $F_{S}(Q,t)$ through the following relation:

\begin{equation}
\displaystyle{F_{S}(Q,t)={1\over N N_{t_{0}}} Re( \sum_{i,t_{0}} e^{i{\bf Q.(r_{i}(t+t_{0})-r_{i}(t_{0}))}}  )          }\label{e1}
\end{equation}
For physical reasons, Q is chosen as the wave vector (here $Q_{0}=2.25$\AA$^{-1}$) corresponding to the maximum of the structure factor $S(Q)$.
$F_{S}(Q_{0},t)$ then allows us to calculate the $\alpha$ relaxation time $\tau_{\alpha}$  of the medium from the equation: 
\begin{equation}
\displaystyle{F_{S}(Q_{0},\tau_{\alpha})=e^{-1}}       \label{e10}     
\end{equation}
Finally, the diffusion coefficient $D$ is obtained from the long time limit of the mean square displacement $<r^{2}(t)>$:
\begin{equation}
\displaystyle{<r^{2}(t)>=   {1\over N N_{t_{0}}}  \sum_{i,t_{0}}  ({\bf r}_{i}(t+t_{0})-{\bf r}_{i}(t_{0}))^{2}                                }       \label{e11}     
\end{equation}
and
\begin{equation}
\displaystyle{\lim_{t \to \infty}  <r^{2}(t)>=  2 d D t                               }       \label{e12}     
\end{equation}

In this work we will chose $\tau_{m}=4 ps$ unless otherwise specified and the dimension $d=3$.


\section{Results and discussion}
 
\subsection{1. Fluidization transition upon activation}

\begin{figure}
\centering
\includegraphics[height=7.2 cm]{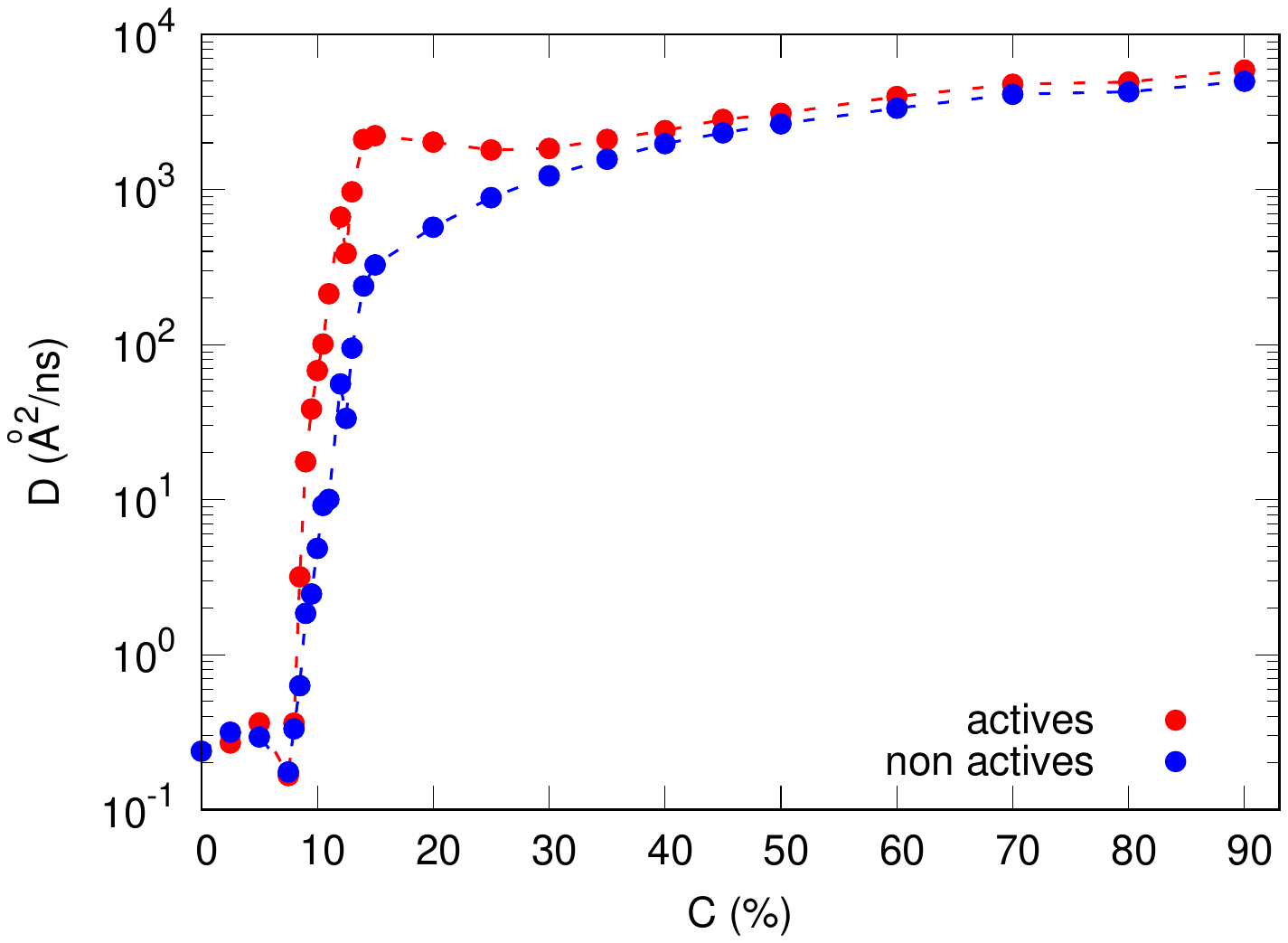}

\caption{(color online)  Diffusion coefficient $D$ versus the concentration $C$ of active molecules at constant temperature $T=220K$. We observe a decrease of $D$ just before the transition, followed by a large increase at the transition. Active molecules are more diffusive than non actives.}
\label{fig1}
\end{figure}

\begin{figure}
\centering
\includegraphics[height=7.2 cm]{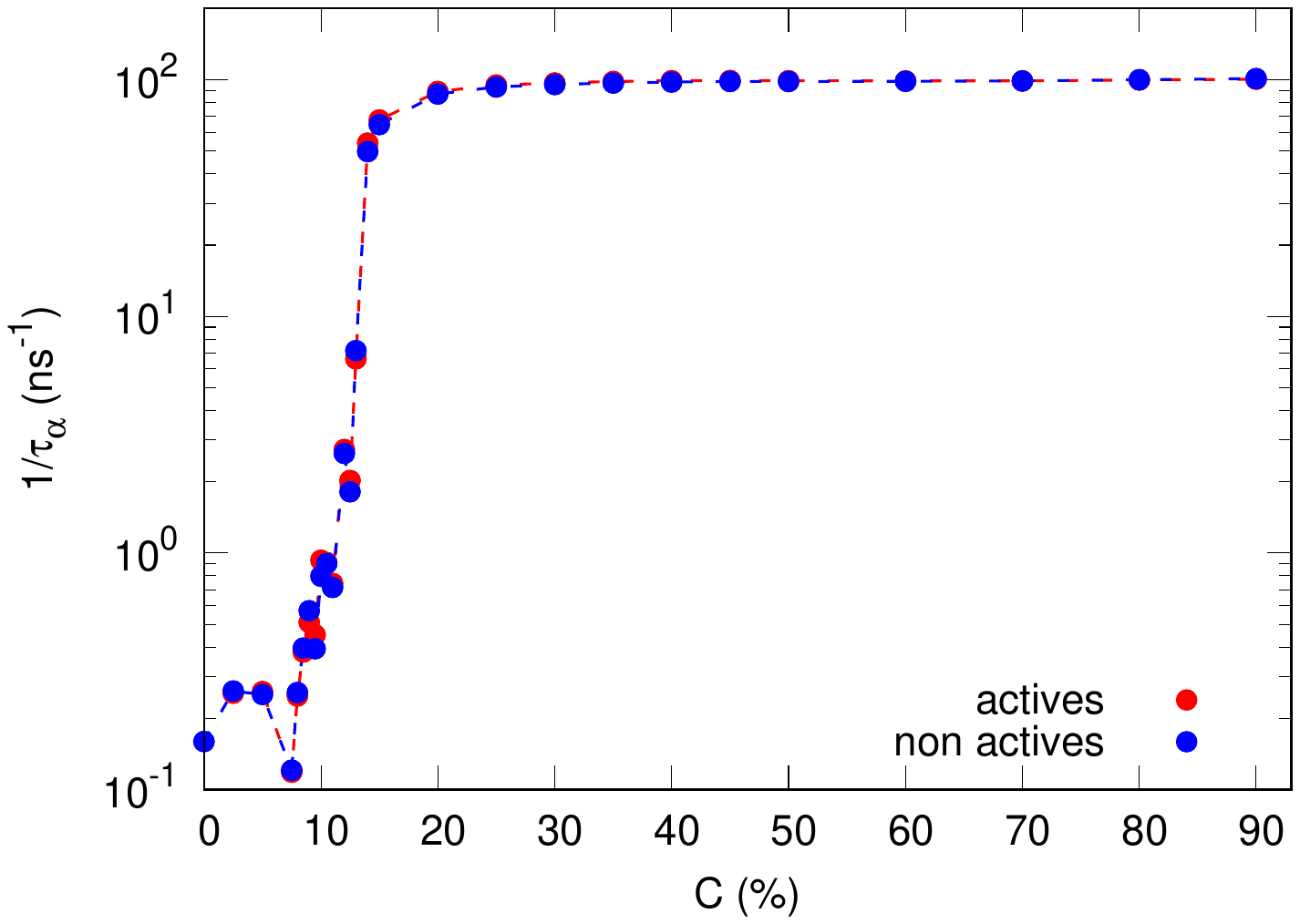}

\caption{(color online)  Inverse of the alpha relaxation time $\tau_{\alpha}$ versus the concentration $C$ of active molecules at constant temperature $T=220K$. The Figure show a decrease of  $1/\tau_{\alpha}$ just before the transition, then a large increase at the transition. Active and inactive molecules display a very similar behavior.}
\label{fig2}
\end{figure}

\begin{figure}
\centering
\includegraphics[height=7.2 cm]{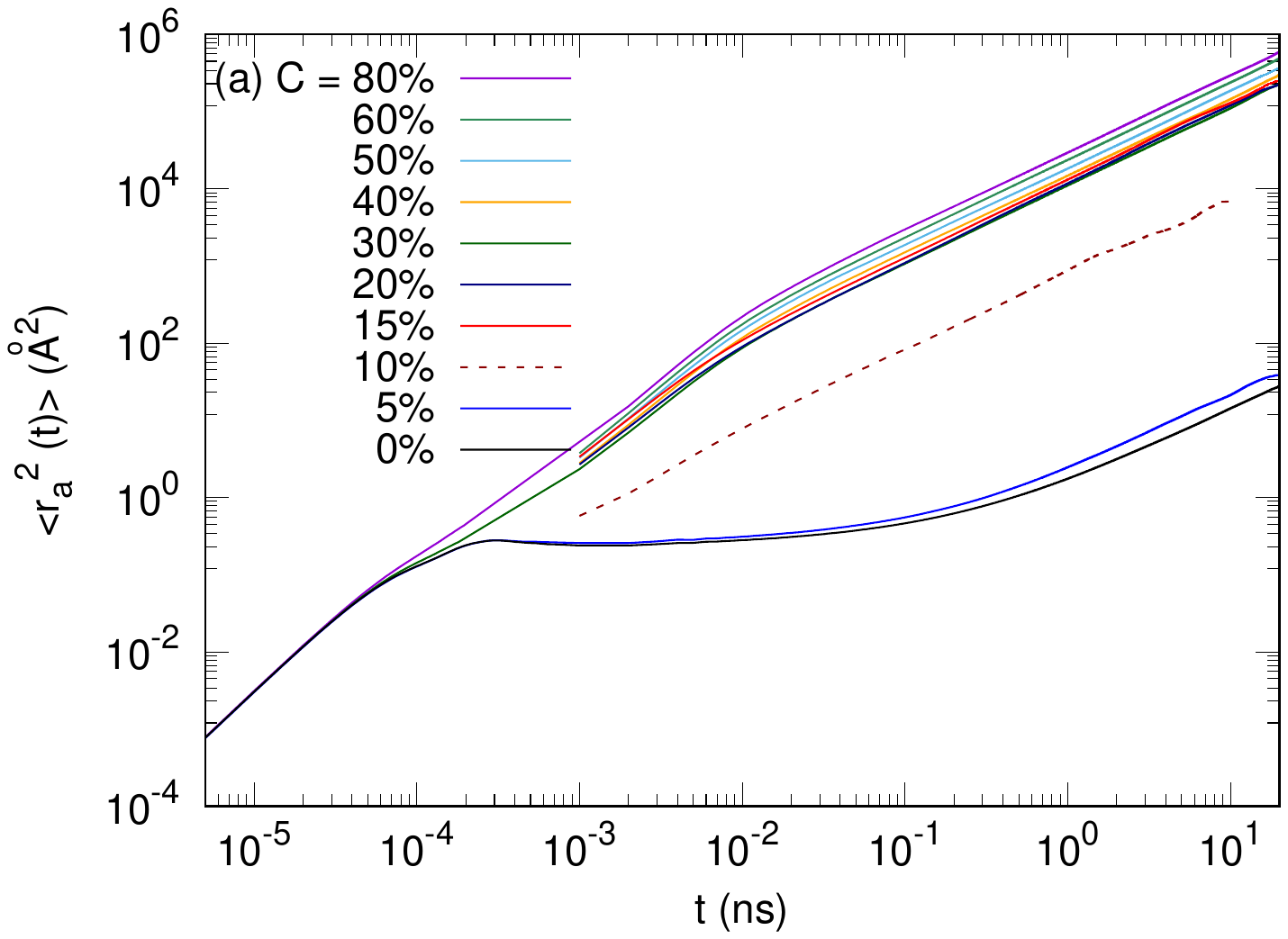}
\includegraphics[height=7.2 cm]{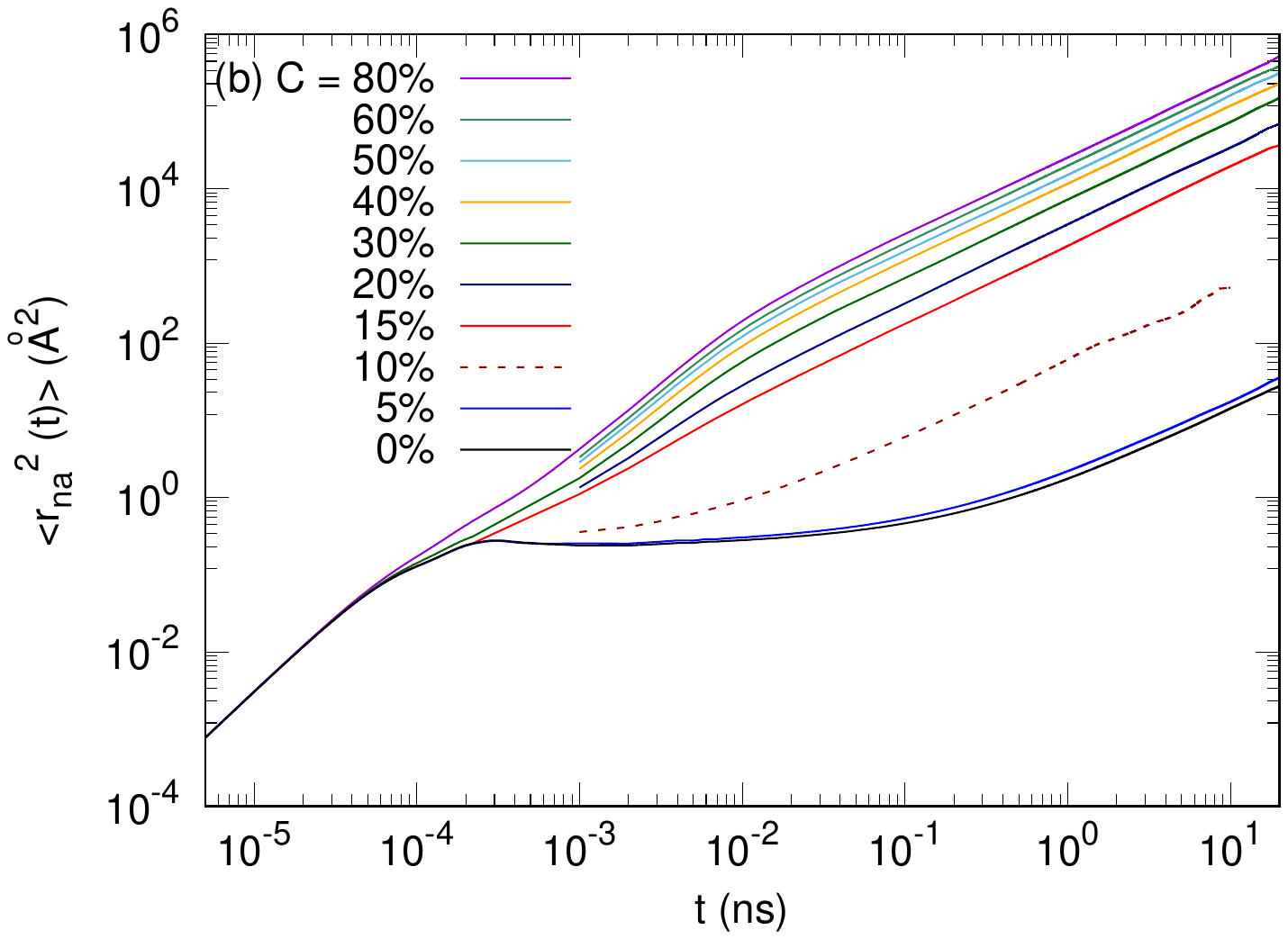}

\caption{(color online)  Mean square displacement of (a) active and (b) non active  water molecules versus the concentration $C$ of active molecules. }
\label{fig3}
\end{figure}

We simulate supercooled water with increasing concentrations of active molecules to probe the transition from a viscous to a fluidized state.
Figures \ref{fig1} and \ref{fig2} show the diffusion coefficient and the $\alpha$ relaxation time $\tau_{\alpha}$ at $T=220\ \mathrm{K}$. A sharp transition occurs near $C_{c}=10\%$. Active and non-active molecules display the same critical behavior in $\tau_{\alpha}$, but the diffusion coefficient jumps more strongly for active molecules. A small decrease in $D$ and $1/\tau_{\alpha}$ appears just before the transition ($C=8\%$), indicating enhanced viscosity with activity. A similar feature was reported in the model system.
Thus, supercooled water exhibits a dynamic phase transition akin to that found in the dumbbell Lennard-Jones liquid, suggesting its universality across supercooled systems.
Figure \ref{fig3} shows the mean square displacement (MSD) of active molecules. Below the transition, the MSD follows the standard three regimes of supercooled liquids: ballistic, plateau, and diffusion. Above the transition, the plateau vanishes, as in liquids above the melting point $T_{m}$, yet cooperative dynamics persist, unlike in equilibrium liquids.
Only $10\%$ active molecules are sufficient to trigger the transition. Since activation occurs only a quarter of the time, the effective fraction is $2.5\%$, while the thermostat holds $T$ constant. The required energy input is therefore far smaller than in a temperature-driven process.
The loss of the plateau indicates rapid cage escape. Because the force exerted by a single active molecule is weaker than the cage-breaking force, several molecules must act cooperatively. We next examine these cooperative motions.

\subsection{2. Cooperative motions above and below the transition}


\begin{figure}
\centering
\includegraphics[height=7.2 cm]{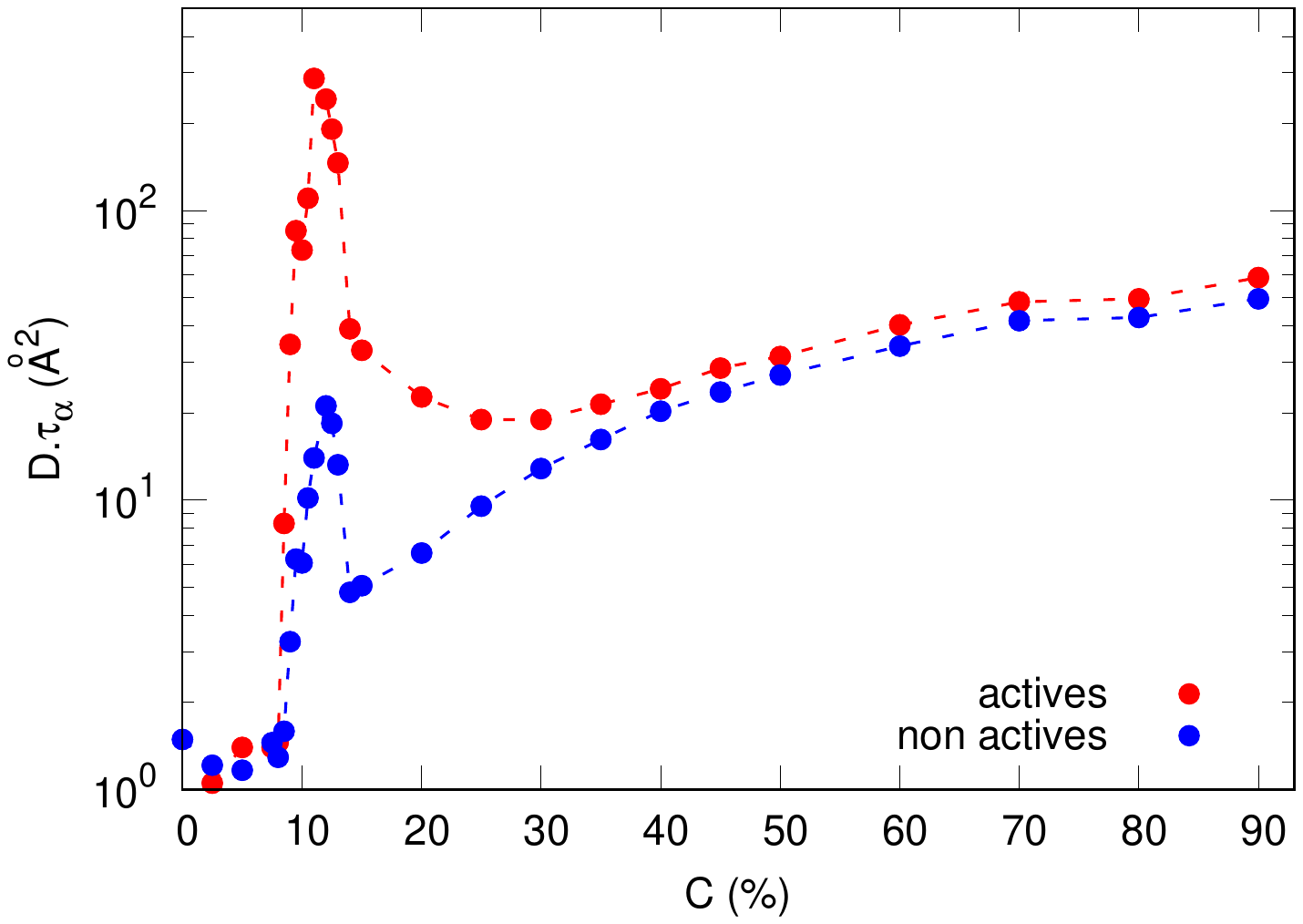}

\caption{(color online)  Stokes-Einstein breaking ($D.\tau_{\alpha}$ variation) versus the concentration $C$ of active molecules at constant temperature $T=220K$. }
\label{fig4}
\end{figure}

\begin{figure}
\centering
\includegraphics[height=7.2 cm]{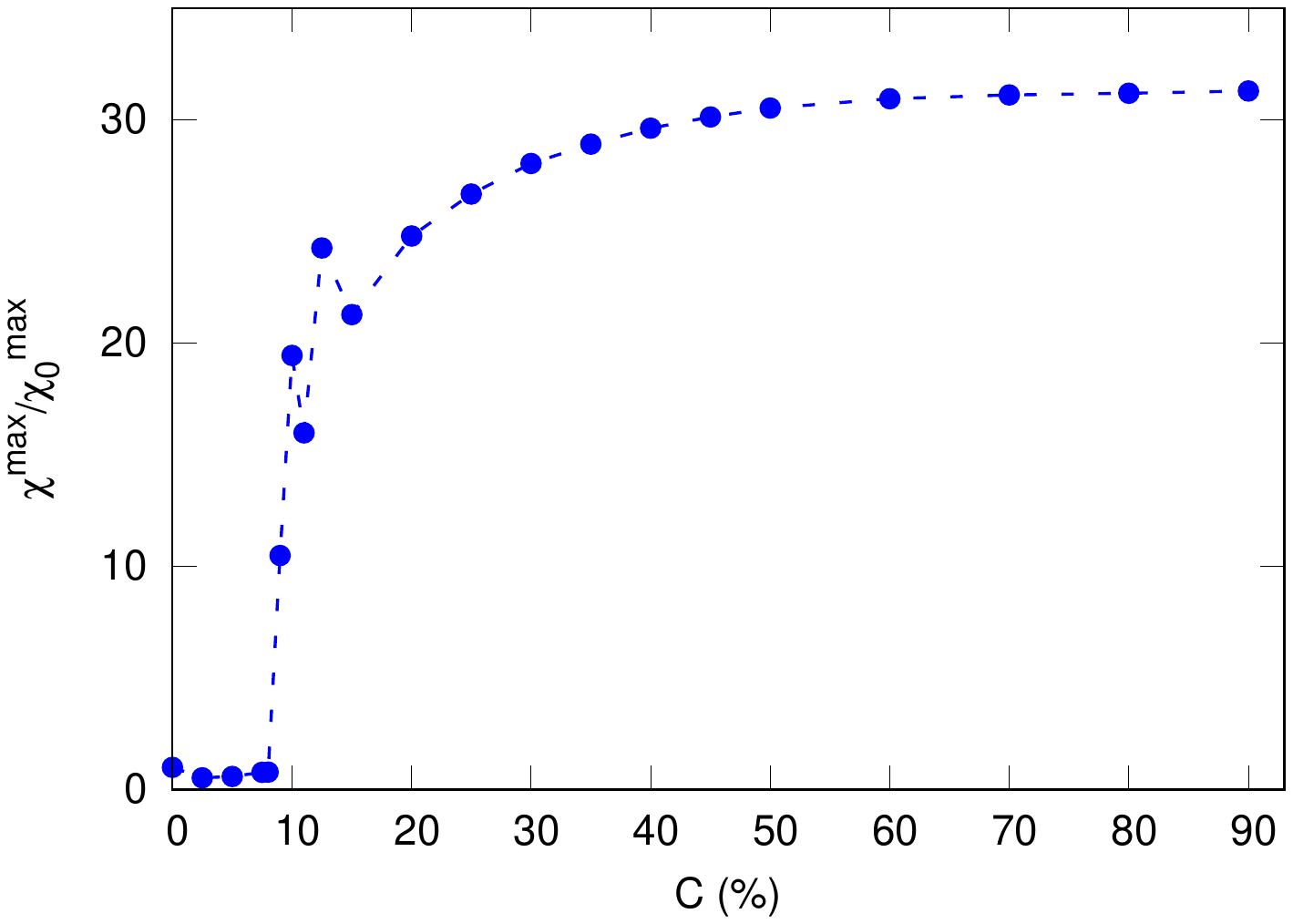}
\caption{(color online)  Maximum value of the dynamic susceptibility of the whole medium versus the concentration $C$  of active molecules.}
\label{fig5}
\end{figure}

\begin{figure}
\centering
\includegraphics[height=7.2 cm]{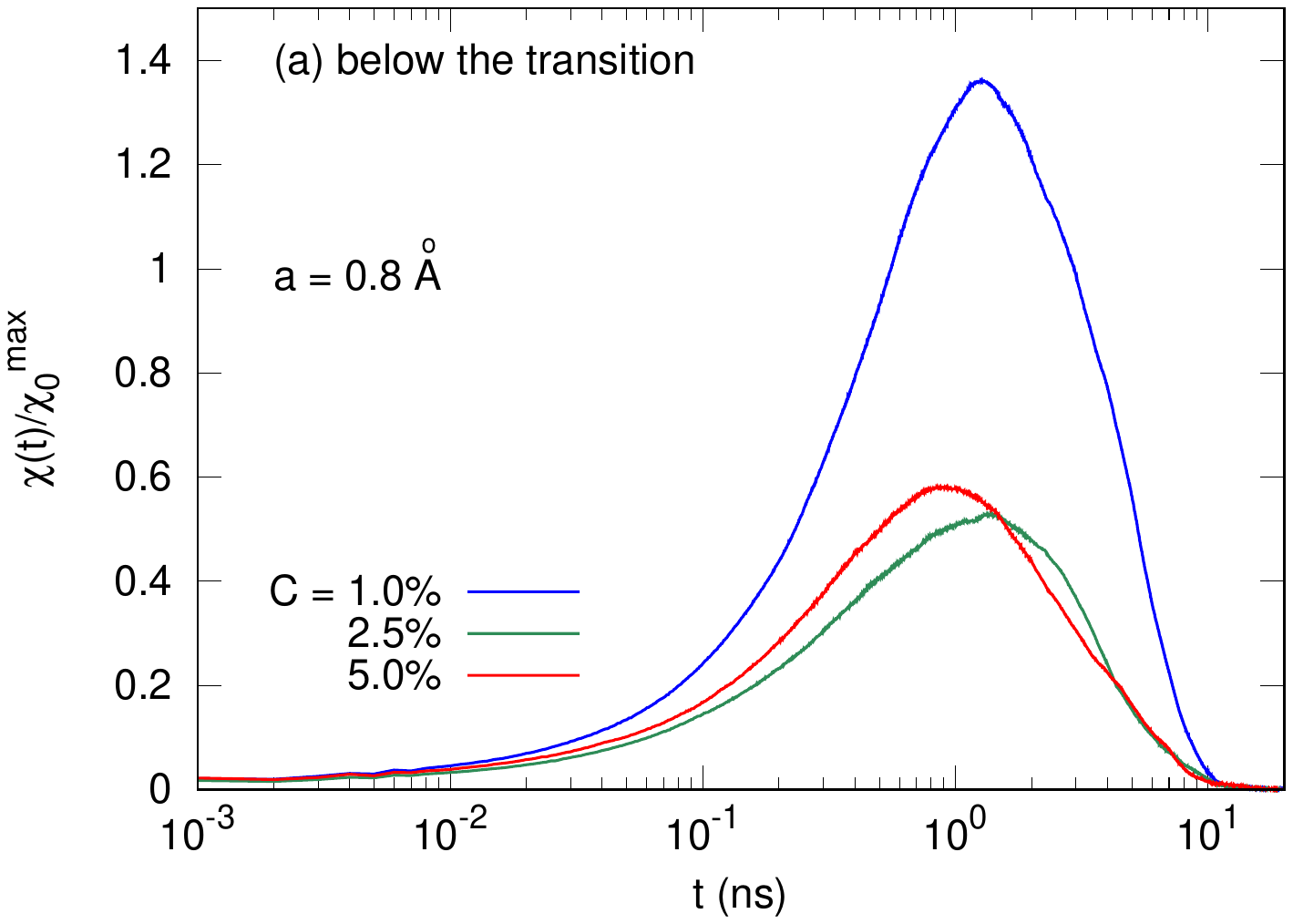}

\caption{(color online) Dynamic susceptibility $\chi_{}(t)$ of the whole medium below the transition, normalized by its maximum value in the non activated liquid $\chi_{0}^{max}$ at the same temperature.}
\label{fig6}
\end{figure}

\begin{figure}
\centering
\includegraphics[height=7.2 cm]{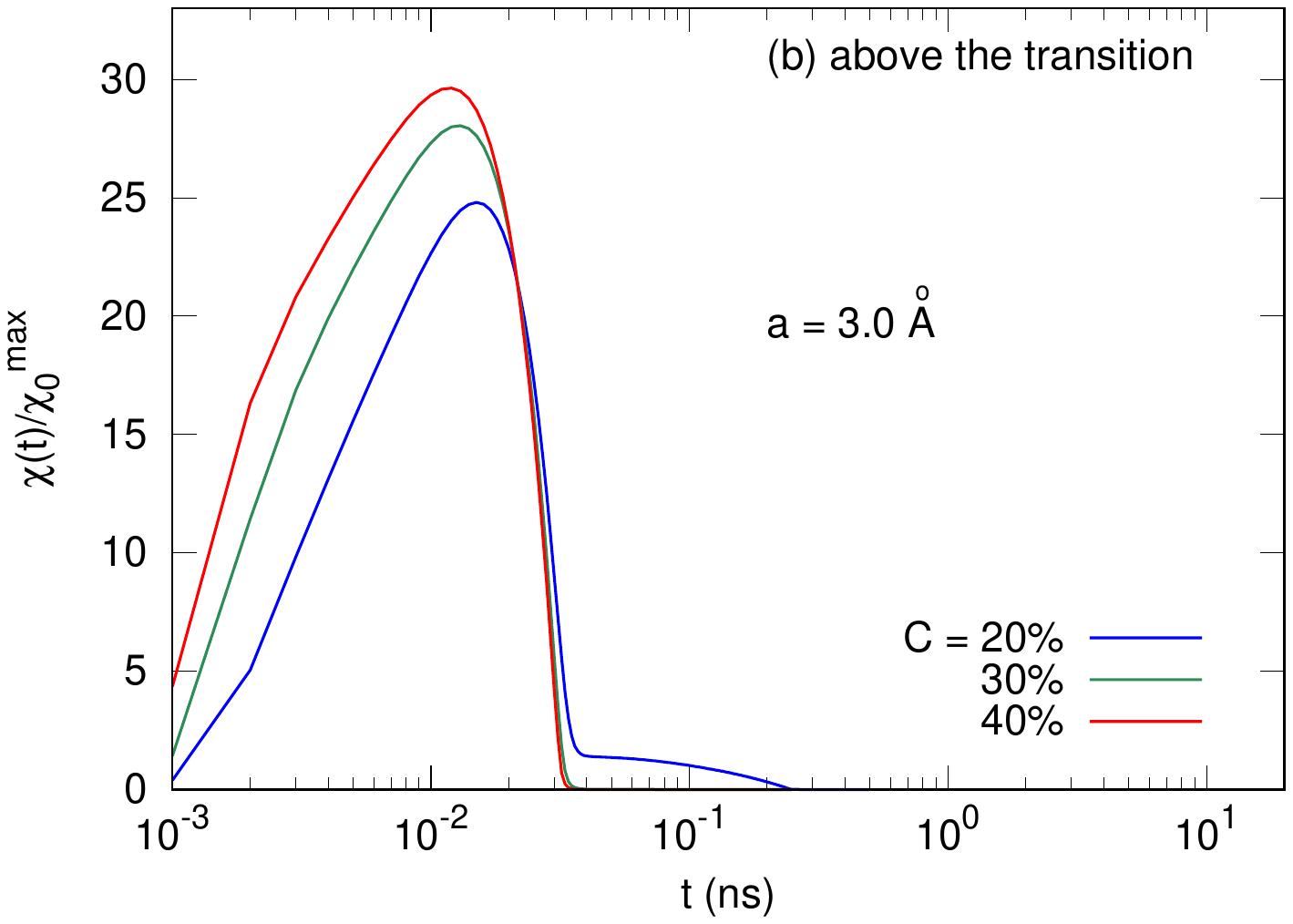}

\caption{(color online)  Dynamic susceptibility $\chi_{}(t)$ of the whole medium above the transition, normalized by its maximum value in the non activated liquid $\chi_{0}^{max}$ at the same temperature.}
\label{fig7}
\end{figure}

\begin{figure}
\centering
\includegraphics[height=7.2 cm]{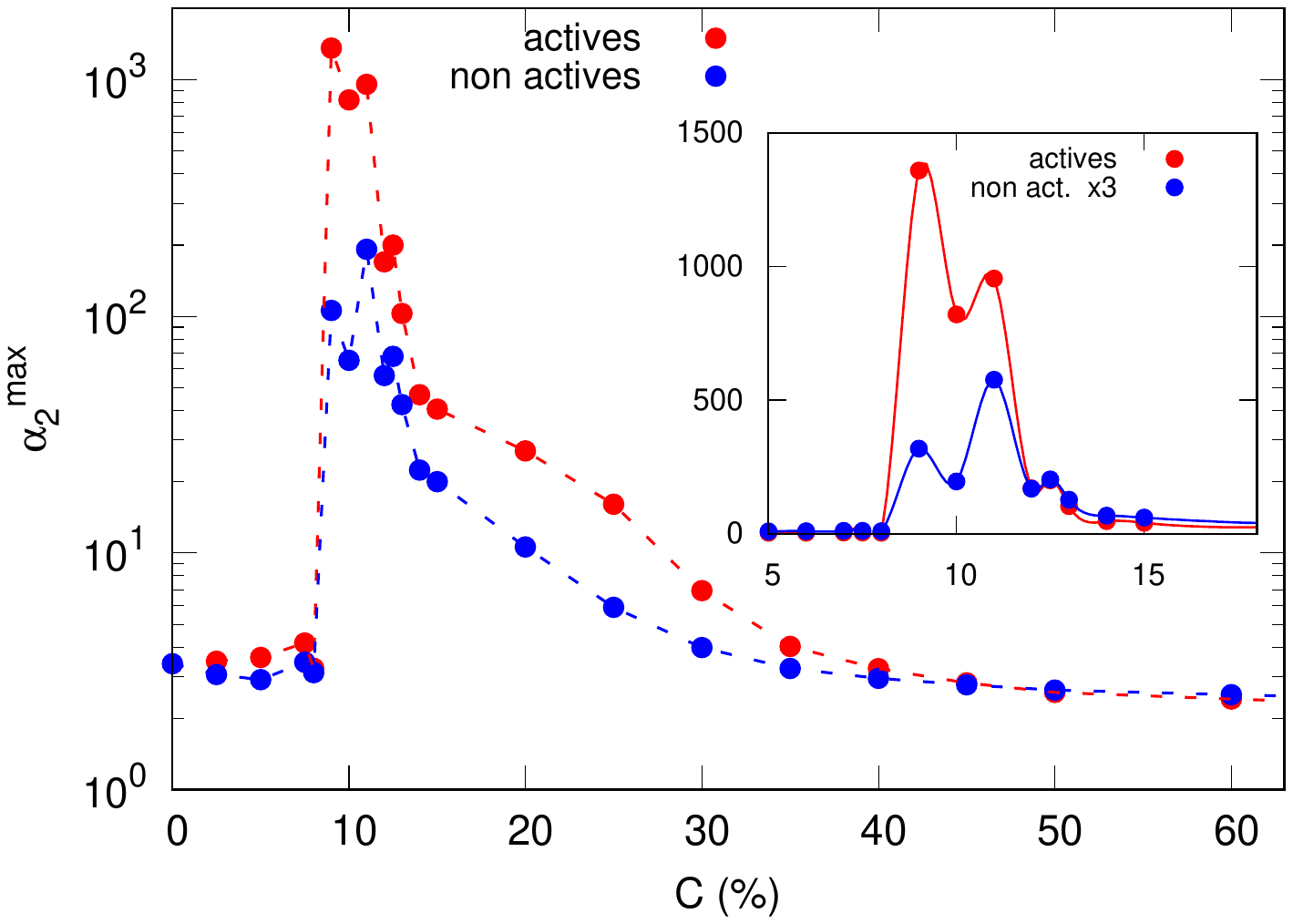}
\caption{(color online)  Maximum value of the non Gaussian parameter $\alpha_{2}(t)$ versus the concentration of active molecules.}
\label{fig8}
\end{figure}

\begin{figure}
\centering
\includegraphics[height=7.2 cm]{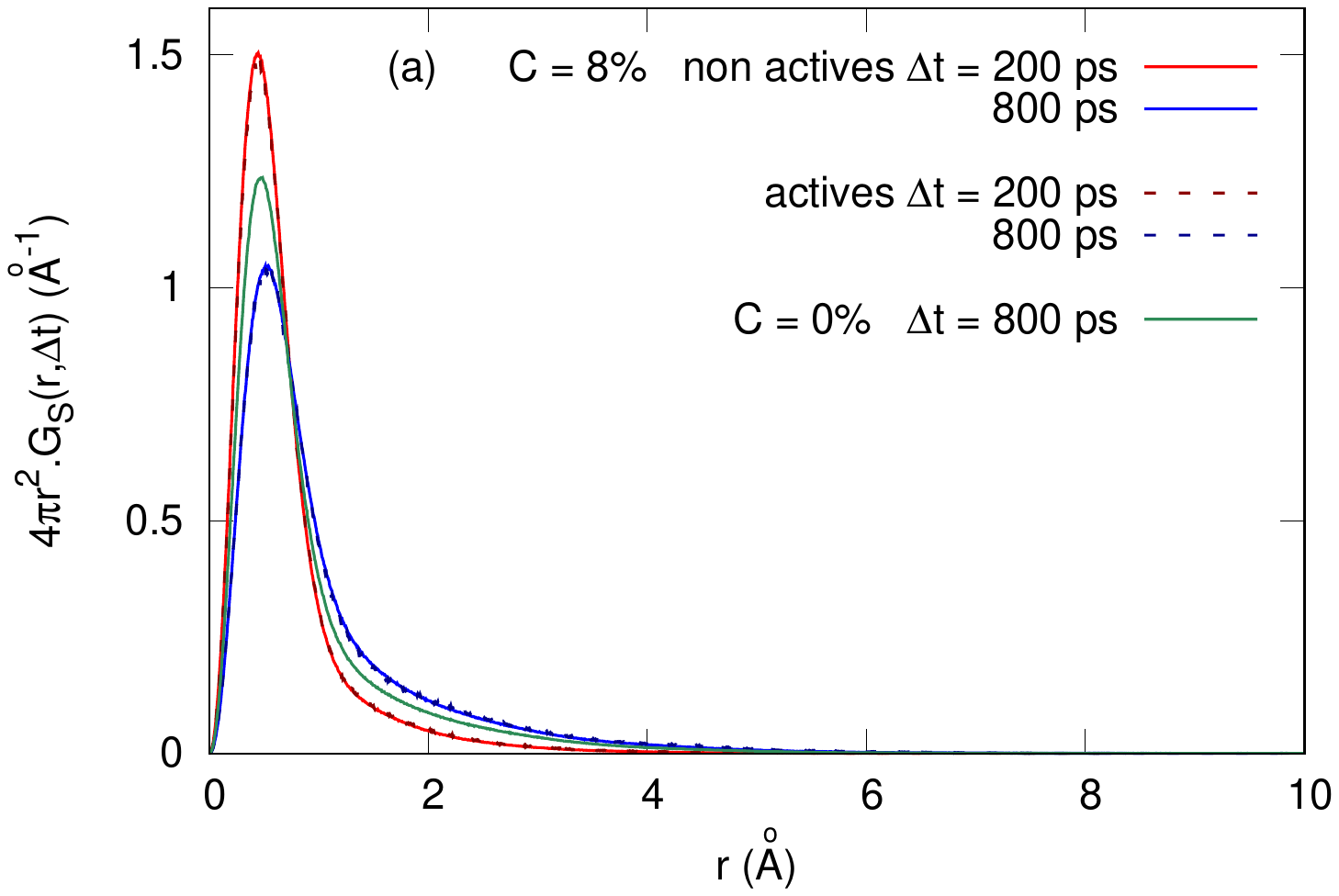}
\includegraphics[height=7.2 cm]{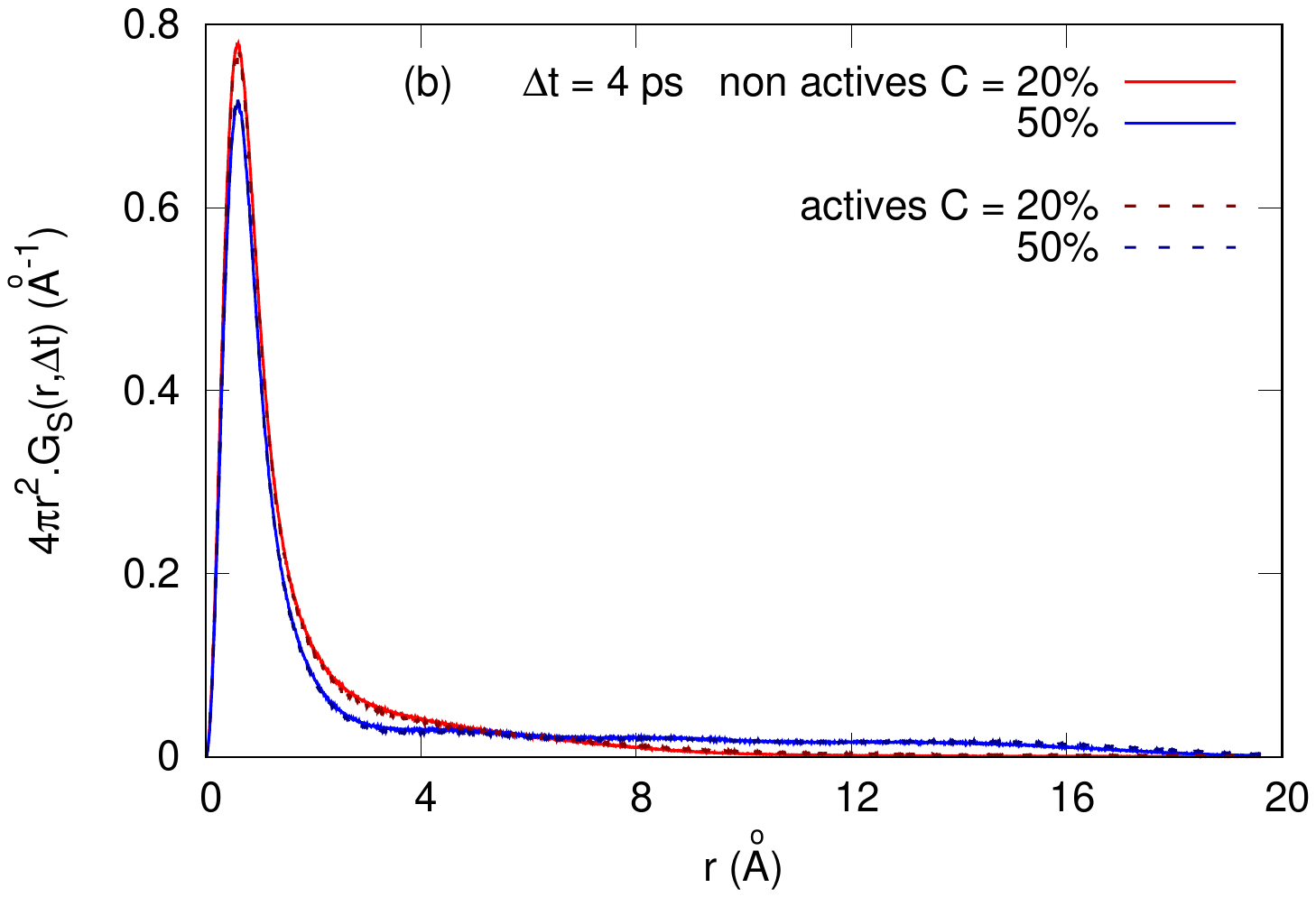}

\caption{(color online)  Self part of the Van Hove correlation function $G_{s}(r,\Delta t)$ (a) below ($C=8\%$)  and (b) above ($C=20\%$ and $50\%$) the transition for different time lapses $\Delta t$ and concentration of active molecules $C$. We observe above the transition large tails, corresponding to displacements ranging from $3$ to $20$ \AA. These tails are typical of cooperative motions in supercooled liquids but are particularly large here. This result suggests that the large increase of diffusion above the transition is induced by cooperative motions.}
\label{fig11}
\end{figure}

In Figures \ref{fig4} to \ref{fig8}  we observe a strong enhancement of cooperative motions above the transition, together with a pronounced peak at the critical concentration. Since such an increase is expected at a phase transition, and since the cooperativity has the same character as in supercooled liquids, we identify it as the physically relevant mechanism driving the transition.
Figure \ref{fig4} shows the evolution of $D.\tau_{\alpha}$ with the concentration of active molecules.
The Stokes–Einstein relation, $D=kT/(6 \pi r \mu)$,
predicts $D\tau_{\alpha} =$ constant at fixed $T$, as $\tau_{\alpha}$ scales with viscosity. In supercooled liquids, however, deviations arise and are attributed to cooperative dynamics. Accordingly, $D\tau_{\alpha}/T$ is widely used as a measure of cooperativity.
We observe in Figure 4 a sharp increase of $D.\tau_{\alpha}$ at the transition, when the concentration of active molecules reaches $C_{c}=10 \%$. That peak is followed by a large increase above the transition,  reaching $\approx 50 $\AA$^{2}$, about $20$ times its value below the transition.
This confirms a strong rise in cooperative motions at the transition. Dynamic susceptibility, the non-Gaussian parameter, and the Van Hove function further support this picture. 
We interpret the peak of cooperative mechanisms at the critical concentration $C_{c}=10 \%$ as the signature of a phase transition\cite{book-phystat1,book-phystat2}.
It implies that the cooperative mechanisms that we observe are the physically relevant ones for the transition.
Non-active molecules also display a weaker peak in $D\tau_{\alpha}$, followed by a rise that merges with the active-molecule values. We interpret this as cooperative motion transmitted from active to non-active molecules, with the transition driven by the active component.

The dynamic susceptibility $\chi$ (Figure \ref{fig5}) shows a similar behavior, a strong enhancement above the transition, with here two peaks at the transition and just above. 
The dynamic  susceptibility $\chi_{}$ is mostly seen as the best measure of  cooperative motions in glass-formers. 
Finding similar results than with the Stokes-Einstein breaking evolution therefore strongly suggests that our observed Stokes-Einstein breaking is due to cooperative motions as in non active glass formers.
Figures \ref{fig6} and \ref{fig7} show  dynamic susceptibilities below and above the transition with an increase by a factor $>30$ above the transition. We observe below the transition (Figure  \ref{fig6}) a decrease of $\chi_{}(t)$ upon activation for small concentrations of active molecules followed by an increase of $\chi_{}(t)$ at larger concentrations. The characteristic time corresponding to the maximum of $\chi_{}(t)$ decreases sharply from $1$ ns below the transition to $10$ ps above the transition,  highlighting the acceleration of cooperative dynamics.
.


The non-Gaussian parameter $\alpha_{2}$ (Figure \ref{fig8}) also shows a structured two-peak maximum at the transition, but decreases at large concentrations, in contrast with $\chi$ and $D\tau_{\alpha}$. Direct analysis of the Van Hove function (Figure \ref{fig11}), however, shows this decrease to be consistent with persistent cooperativity at high $C$. 
The $\alpha_{2}$ peak is larger for active molecules than for non-actives, again confirming that active molecules drive the dynamics. Interestingly, the first peak dominates for active molecules, while the second is stronger for non-actives, possibly reflecting the two structural states (LDL and HDL) of water at this temperature.

To clarify the decrease of $\alpha_{2}$, we examine the Van Hove self-correlation function $G_{s}(r,\Delta t)$ (Figure \ref{fig11}),  which deviation to the Gaussian shape predicted by Brownian motion is measured by the Non Gaussian Parameter.
$G_{s}(r,\Delta t)$ represents the probability distribution for a molecule to be  a distance r apart its initial position after a time lapse $\Delta t$. The existence of a tail in the Van Hove corresponding to molecules moving more than expected from the Gaussian law represent in supercooled liquids molecules moving cooperatively.
We observe that tail when $C=0\%$ due to spontaneous cooperative motions in supercooled liquids. When $C$ increases, below the transition, the tail increases due to active molecules. Notice that active and non active molecules display the exact same behavior, showing that the motion of active molecules is transmitted to non-actives. Above the transition the tail increases largely, showing that some molecules move much farther than expected.
These results show that fluidization, i.e. the increase of diffusion, is directly linked to enhanced cooperative dynamics.


\begin{figure}
\centering
\includegraphics[height=7.2 cm]{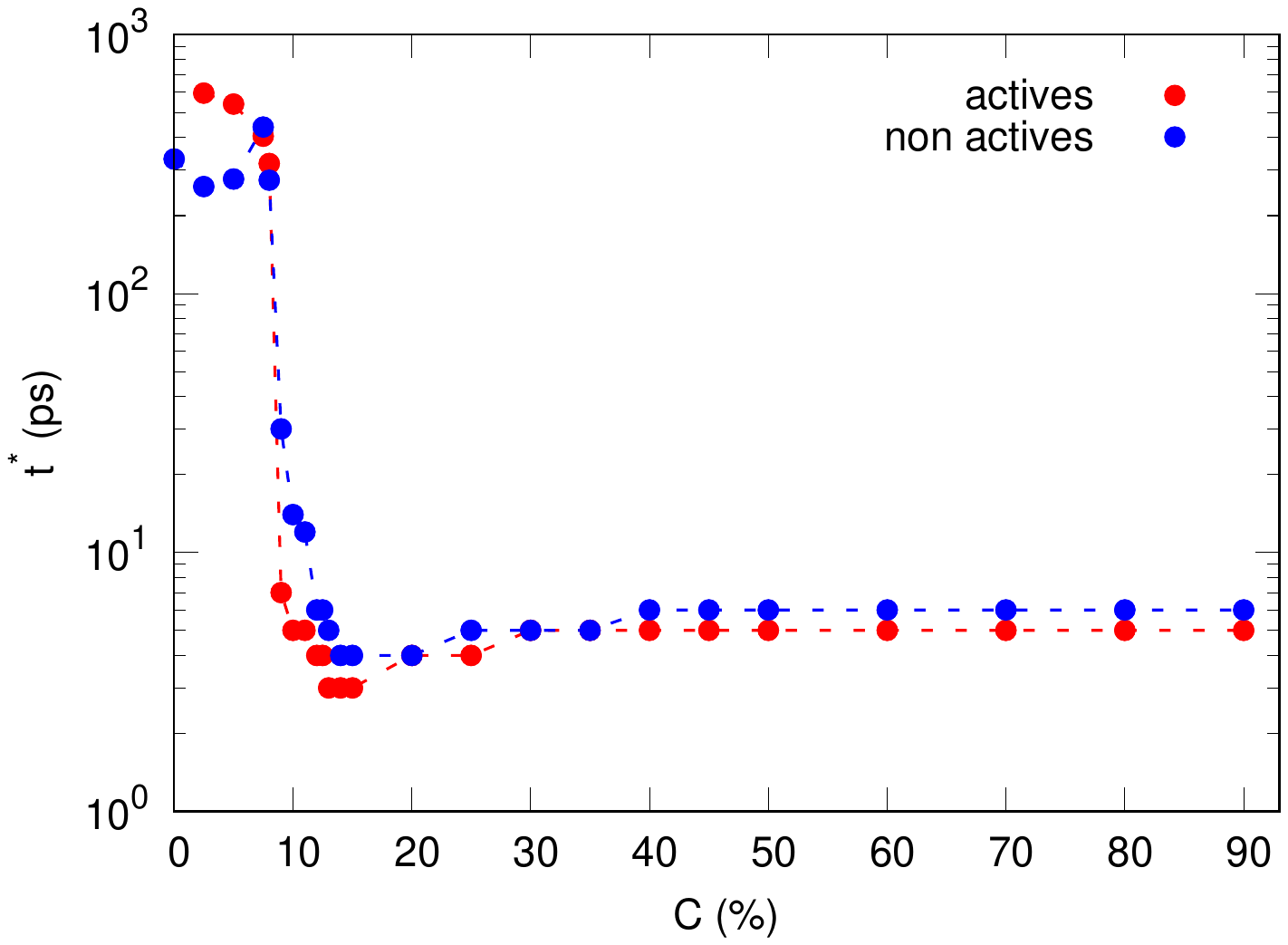}

\caption{(color online)  Time $t^{*}$ corresponding to the maximum value of the Non Gaussian parameter $\alpha_{2}(t^{*})=\alpha_{2}^{max}$. In supercooled liquids $t^{*}$ is known as the characteristic time of dynamic heterogeneities (that is of cooperative motions).}
\label{fig9}
\end{figure}

\begin{figure}
\centering
\includegraphics[height=7.2 cm]{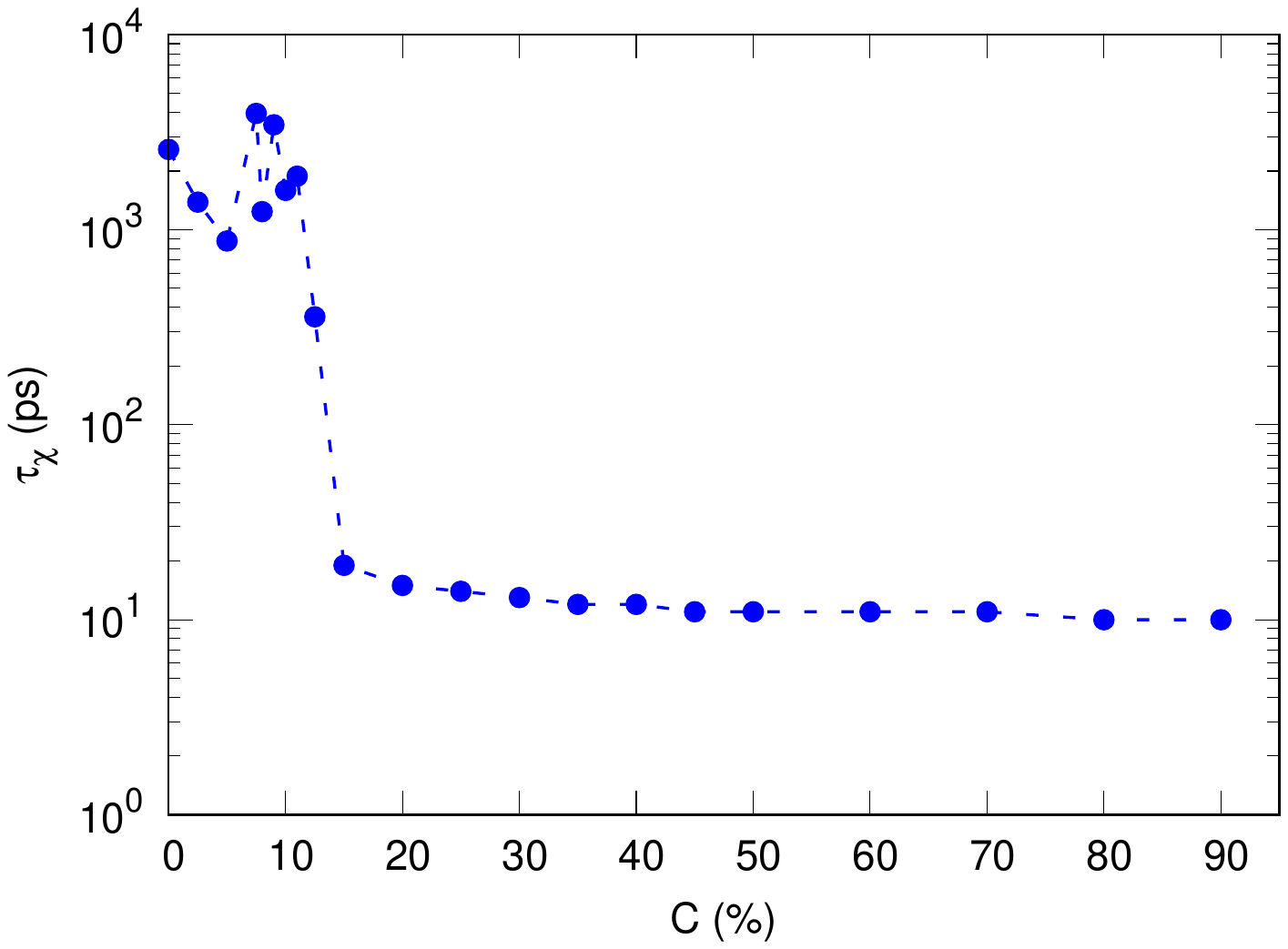}

\caption{(color online)  Characteristic time $\tau_{\chi}$ of the dynamic susceptibility $\chi$.}
\label{fig10}
\end{figure}

\vskip0.5cm
{\color{Black} Figures \ref{fig9} and \ref{fig10} show the characteristic times of $\alpha_{2}$ and $\chi$ as functions of concentration. Both drop sharply at the transition, indicating that cooperative mechanisms shift to much shorter timescales above $C_{c}$. This acceleration of cooperative motions, driven by periodic activation, reduces relaxation times and underlies the observed fluidization.}

\subsection{4. Structural modifications at the transition}

\begin{figure}
\centering
\includegraphics[height=7.2 cm]{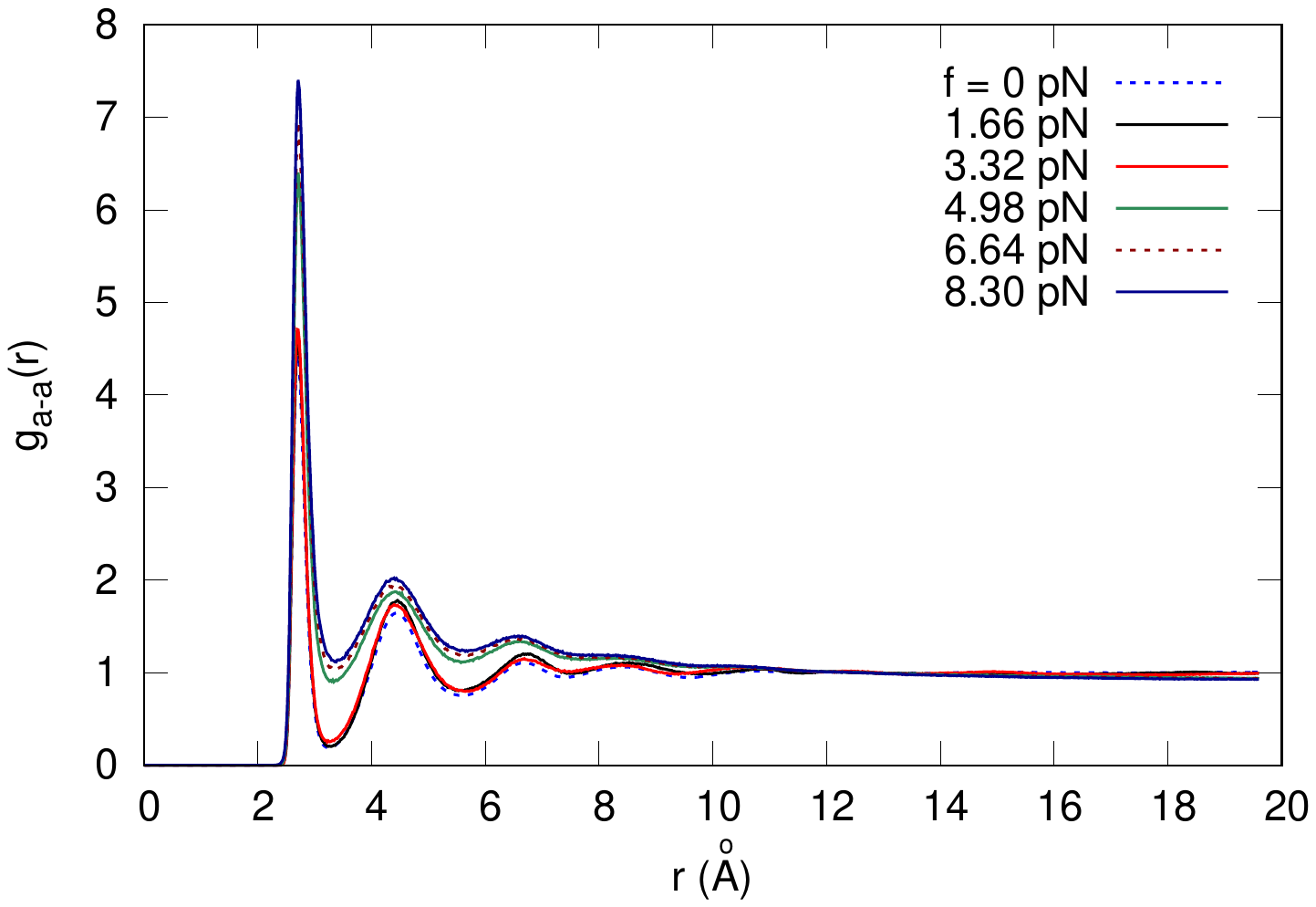}

\caption{(color online)  Radial distribution function between active water molecules for various values of the activation force, for a concentration $C=10\%$.}
\label{fig12}
\end{figure}

\begin{figure}
\centering
\includegraphics[height=7.2 cm]{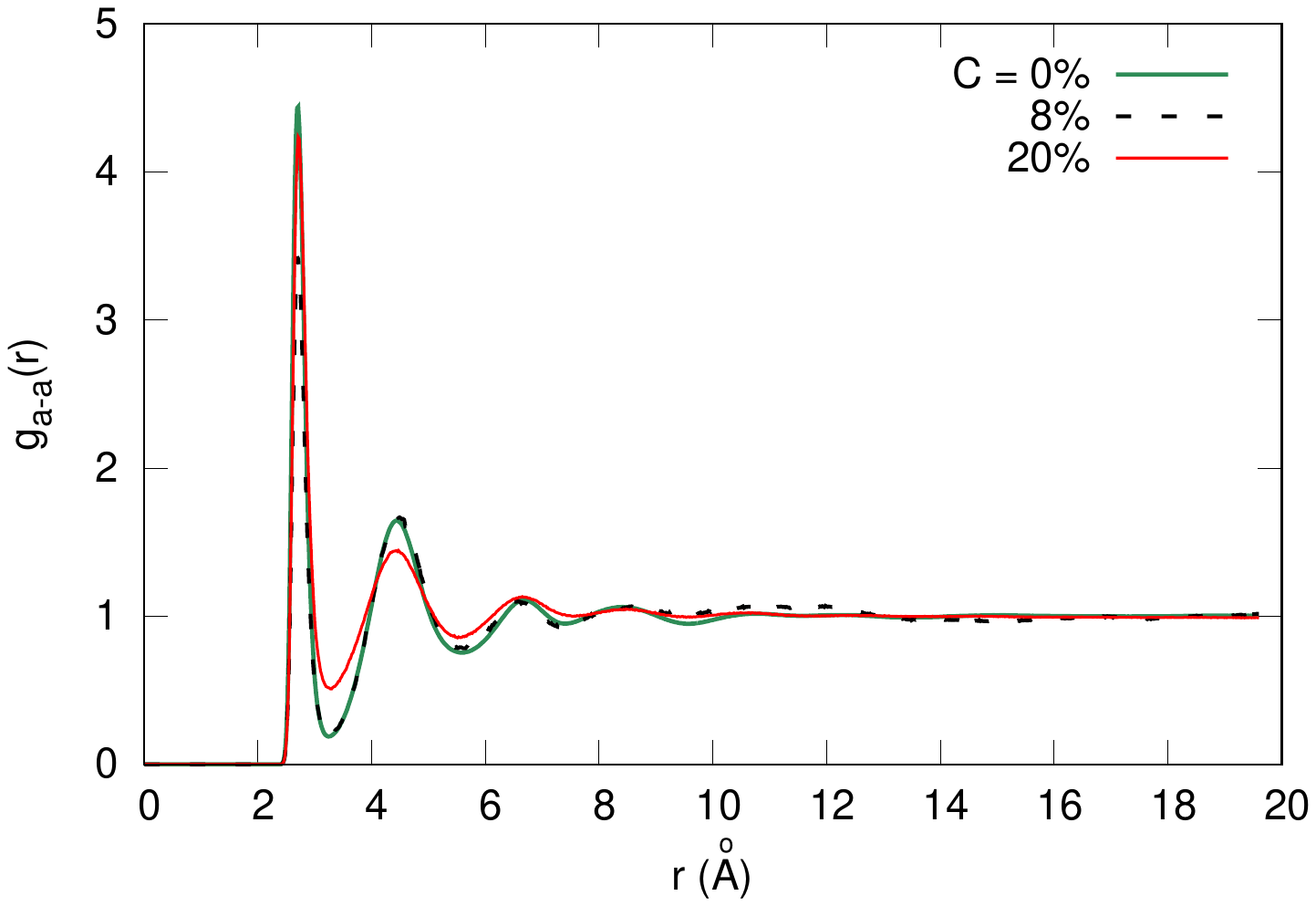}

\caption{(color online)  Radial distribution function between active water molecules just below and above the transition.}
\label{fig13}
\end{figure}

\begin{figure}
\centering
\includegraphics[height=7.2 cm]{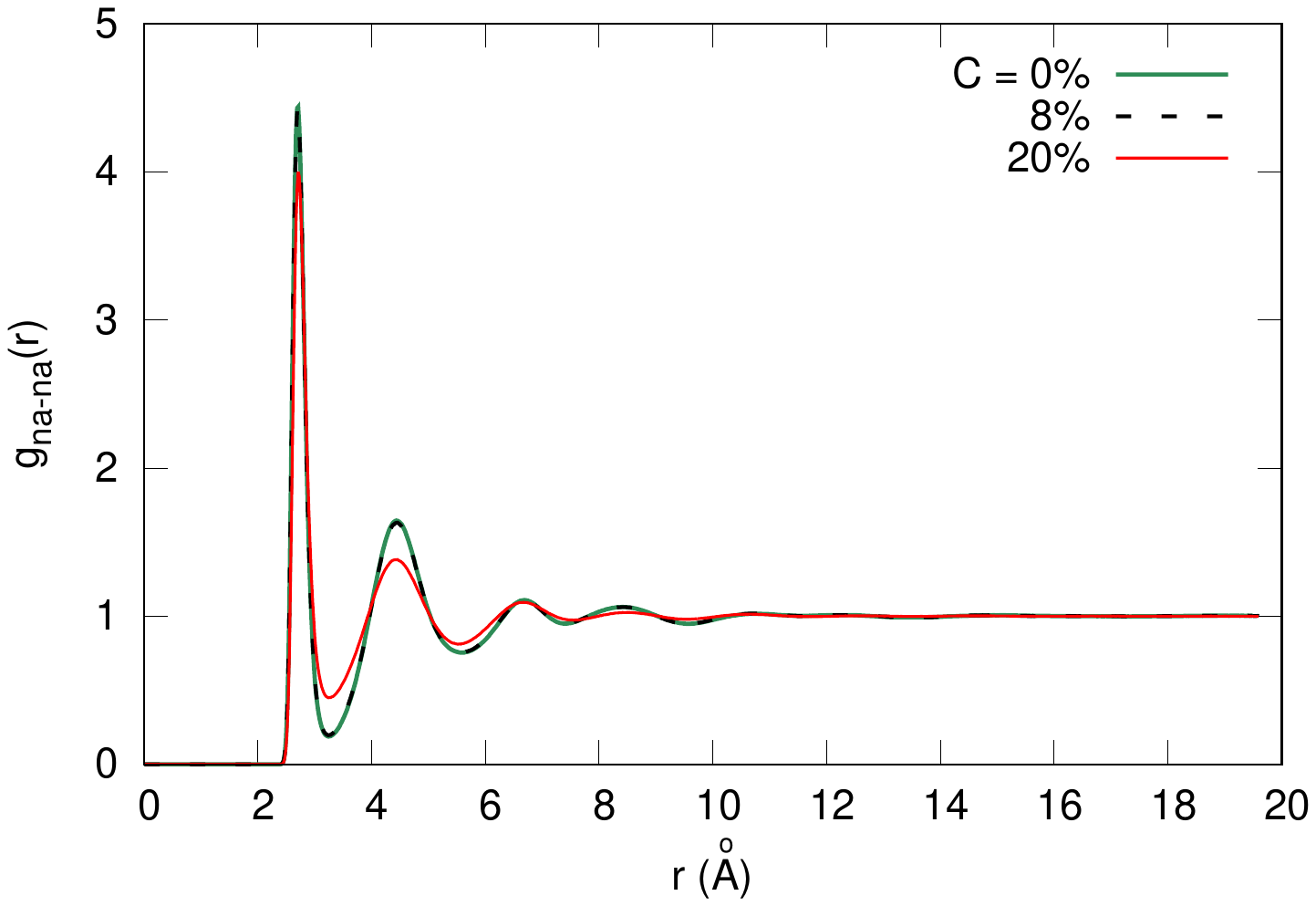}

\caption{(color online)  Radial distribution function between non active water molecules just below and above the transition.}
\label{fig14}
\end{figure}

Active water molecules do not show a clear aggregation in Figure \ref{fig13} ($f=3.32$ pN) but we observe a change in structure that is consistent with an LDL to HDL transition, at the transition.
Non active water molecules display the same structural change in Figure \ref{fig14} that seems to be mixed with a small aggregation.
 Figure \ref{fig12} shows the evolution of the radial distribution function between active molecules with the activation force.
We observe an aggregation of the active molecules when the force $f$ is larger than the threshold  $f_{c}\approx4$ pN.
We then find an aggregation of active molecules as in the previous work with a model liquid.
However the transition appears below that threshold as most of our study corresponds to $f=3.32$ pN.
We expect however an aggregation of the most mobile molecules due to the presence of dynamic heterogeneity.

\section{Conclusion}

Using an activation mechanism mimicking facilitation, we recently found a dynamic phase transition induced by a small fraction of active molecules in a model system. 
Motivated by this finding, we investigated in this work whether a similar transition occurs in supercooled water. We found clear evidence of the transition despite water’s numerous anomalies, supporting the universality of the phenomenon. At $220,\text{K}$, only $2.5\%$ of active molecules are sufficient to trigger fluidization in water, with the required fraction decreasing at higher temperatures. 
As in the model system, we observe in supercooled water a strong increase in cooperative mechanisms at the transition.
Such cooperative behavior is a usual signature of phase transitions\cite{book-phystat1,book-phystat2}.
Importantly, these mechanisms coincide with the spontaneous cooperative dynamics typically observed in supercooled liquids, suggesting that they are the relevant ones driving both the fluidization and the glass transition.
In contrast to previous studies, we do not observe a clear aggregation of active molecules at the transition in water. We interpret this absence as the result of an interplay with a concurrent transition between LDL and HDL, also triggered by activation.
This additional transition is particularly significant in the context of preventing water crystallization, since LDL is a known precursor to crystallization, and work is in progress to better specify that transition.

 }









\end{document}